# Easter Island: the Tongariki and Mataveri Solar Observatories Used a Common Methodology

Sergei Rjabchikov[1]

[1]The Sergei Rjabchikov Foundation - Research Centre for Studies of Ancient Civilisations and Cultures, Krasnodar, Russia, e-mail: srjabchikov@hotmail.com

**Abstract**

Two additional positions of the famous Mataveri calendar of Easter Island have been interpreted. The information about the stars of the Virgo constellation has been obtained, too. As a result, the archaic Rapanui name of the star Spica and the Hawaiian name of this star have been compared.

## Introduction

The great civilisation of Easter Island is famous in different aspects (magnificent statues and platforms, secret rites, and some objects covered with mysterious signs lastly). Here we continue the study of the archaeoastronomy of this ancient society.

## On the Astronomical Observations at Tongariki in the Past

In conformity with Mulloy, the ceremonial platform (*ahu*) Tongariki was a real solar observatory (Liller 1991: 270). It is no doubt that the peculiar information about this is retained in the local folklore. Barthel (1978: 82) has found a passage in Manuscript E about this *ahu*:

*ko tongariki a henga eha tunu kioe*
*hakaputiti ai ka hakapunenenene*
*henua ma opoopo o tau kioe*

otherwise,

*Ko Tongariki: a henga e ha tunu kioe.*
*Hakapu titi ai, ka hakapu nenenene*
*henua ma opoopo o tau kioe.*

(This is) *Tongariki* [*Tonga Ariki*]: during the fourth dawn (*henga*) a rat is cooked (as a sacrifice).
Lift the instrument (*titi*) at the spot, lift (it) in the same time interval
at the land for the careful study during the season of the rat (= the bright sun).
(*This is the translation of mine*.)

The text is full of the archaic terminology: cf. Mangarevan *pupu* ‘to grow,’ *nenea* ‘to abound; to multiply’ and *opo* ‘to inspect; to pay attention.’ It is clear that in the fourth morning starting with the night of the new moon (before the time of a possible solar eclipse) the dimensions of the moon were measured. Why did the ancient priest-astronomers keep watch on the moon in this manner? The fourth crescent was well visible in the sky, and knowing its proportions, one could decide that the invisible new moon several nights ago had been predicted (counted) correctly indeed. The eighth morning (the number four repeated twice) could be chosen to measure the dimensions of the moon, too.

## Further Remarks about a Rapanui Rock Calendar

On a boulder at Mataveri (a centre of the bird-man cult) some lines were incised; most of them were the directions of the setting sun in compliance with Liller (1989). I have determined the corresponding days for the year 1775 A.D. (Rjabchikov 2014: 5, table 2). From the interpretation offered above it is obvious that the native astronomers at Mataveri used two dates, September 27 and October 1, to reveal that the moon had been first new exactly on September 24, see table 1.

**Table 1.** The Dates Calculated (with the interpretations for September 27 and October 1):

**June 22** (the azimuth of the sun = 296.2°): one day after the winter solstice;
July 21 (292.5°): the star Capella (α Aurigae) before dawn;
August 11 (286.7°): the star Pollux (β Geminorum) before dawn;
September 2 or 3 (277.9°): the star β Centauri [*Nga Vaka*] before dawn;
**September 21** (270.1°): the day before the vernal equinox, the key moment of the bird-man feast;
September 24 (268.7°): the new moon;
September 27 (267.4°); **the fourth night: the measure of the visible dimensions of the moon;**
October 1 (265.9°); **the eighth night: the measure of the visible dimensions of the moon;**
October 3 (264.7°);
October 22 (256.8°): near the new moon;
November 8 (250.7°): the star Spica (α Virginis) before dawn;
November 12 (249.3°);
November 14 (248.7°);
November 23 (246.3°): the new moon;
**December 20** (the azimuth of Aldebaran = 339.1°): the star Aldebaran (α Tauri) at night;
**December 21** (the azimuth of Aldebaran = 322.1°; the azimuth of Canopus = 177.5°): the stars Aldebaran (α Tauri) and Canopus (α Carinae) on the same night (Rjabchikov 2013: 7); the day of the summer solstice.

Consider the parallel text on the Aruku-Kurenga tablet (B), see figure 1.

1(Bv 4): 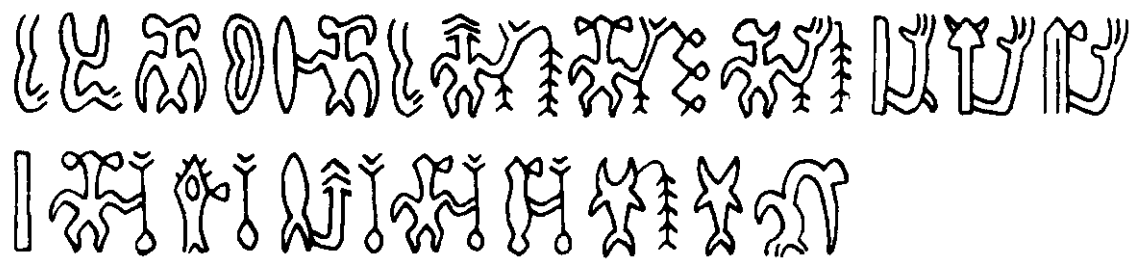

Figure 1.

1 (Bv 4): **43 2 44 47 30-44 43 33 6-15 24 6-15 52 6-15 24 4-15 21-15 26-15 4-6 101 56 101 12 4-33 101 6 101 56 101 11 24 11 19** *Ma Hina Ta(h)a avae Anakena, ma ua, Hora ari, Hora iti, Hora ari. Atua roa Koro Maro tuha. O(h)o po, o(h)o Ika Atua, o(h)o Ha, o(h)o po, o(h)o Mango ari, Mango ki.* 'The moon (*Rongo*) *Tane* of the month *Anakena* [July chiefly] goes, the rains, (then the month) *Hora* [*Hora-Nui*; September chiefly] of the bright sun after (the month) *Hora-Iti* [August chiefly], (i.e. the month) *Hora* [*Hora-Nui*; September chiefly] of the bright sun go. The great god (the month) *Koro* [December chiefly] (came) (after) the time interval (month) *Maro* [June chiefly]. The nights enter, the moon *Atua* enters, the constellation Auriga enters, the nights enter, the star Castor (?) or Pollux (?), or the star Regulus (?), or star Spica (?) enters.'

To understand the meaning of the *rongorongo* expression *Mango ari*, *Mango ki* with two "SHARK" signs examine a fragment inscribed on the Berlin tablet (O) which was a textbook in the royal *rongorongo* school of king *Nga Ara* (Rjabchikov 2012: 22), see figure 2.

1(O3[5]): 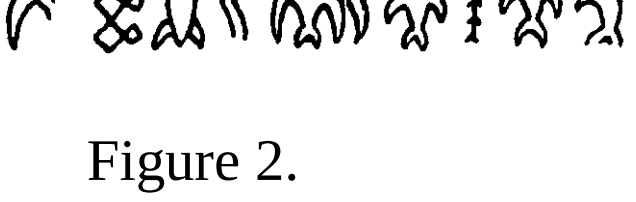

Figure 2.

1 (O3[5]): **2** (a reversed variant) **17 28-15 44 11 24 11 19** *Ina tea, ngaro. Taha Mango ari, Mango ki.* '(It is) the lack of whiteness, (the sun is) eclipsed. (The star) *Mango ari, Mango ki* is moving.'

The solar eclipse of September 16, 1773 A.D. was the last total till the collapse of the Rapanui culture because of the Peruvian raid brought the majority of the natives to slavery in 1862 A.D. King *Nga Ara* could order to copy the *rongorongo* text about the eclipse from a tablet belonged earlier to his father, king *Kai Makai* the First. Here and everywhere else, I use the computer program RedShift Multimedia Astronomy (Maris Multimedia, San Rafael, USA) to look at the heavens above Easter Island.

The sun was in the Virgo constellation during that total eclipse. *Mango ari, Mango ki* could be either the name of this constellation or the name of its bright star Spica (α Virginis) or even Spica and another star in the constellation. Here Old Rapanui *ari* means 'bright, visible,' and *ki* means 'full.'

In the Hawaiian native astronomy, the term *Hikianalia*, by the bye, denotes Spica (Johnson and Mahelona 1975: 3). In my opinion, the name reads *Hiki ana alia* [*Hiti ana aria*] 'The appearance of much of brightness.' So, in both Rapanui and Hawaiian names of Spica is the common term, *ari = ali(a)* 'bright; brightness.'

**On the Observations of Spica Discussed in Rapanui Parallel Records**

Consider then three parallel records on the Great St. Petersburg (P), Great Santiago (H) and Small St. Petersburg (Q) tablets, see figure 3.

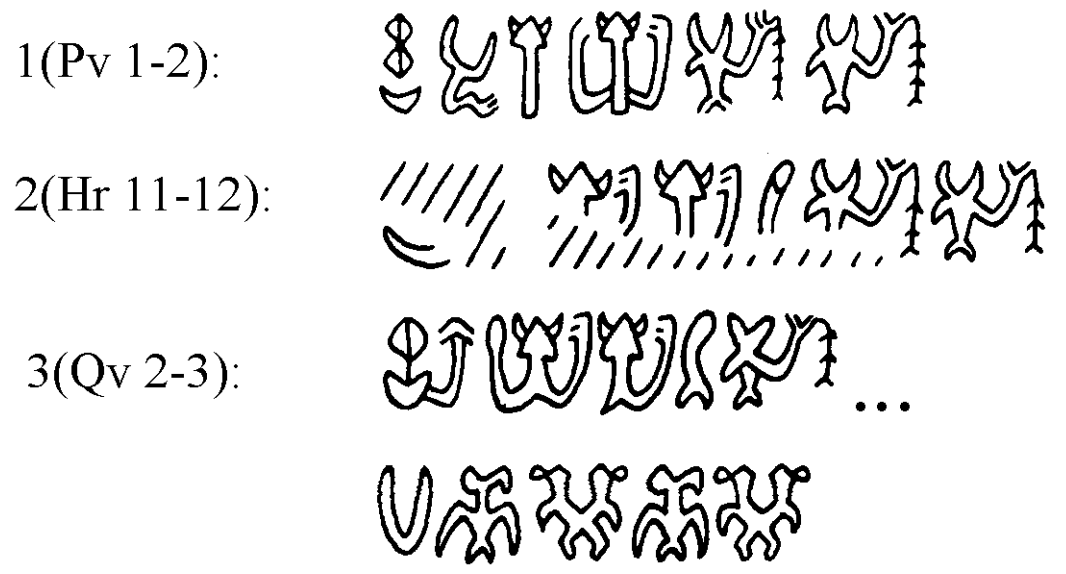

Figure 3.

1 (Pv 1): **3 17 2 21-15-21-15 11 24 11 24** *Hina* (= *marama*) *tea Hina Korokoro; Mango ari, Mango ari.* '(It is) the white moon of the moon goddess of the month *Koro* (*Korokoro* literally; December chiefly); (it is the star) *Mango ari, Mango ari*.'

2 (Hr 11-12): **3** [**17**] **21-15-21-15 12 11 24 11 24** *Hina* (= *marama*) [*tea*] *Korokoro; IKA Mango Ari, Mango Ari.* '(It is) the [white] moon of the month *Koro* (*Korokoro* literally; December chiefly); (it is the star) FISH *Mango ari, Mango ari*.'

3 (Qv 2): **3 17 4-33 4 21-15-21-15 12 11 24 … 61 44 6var 44 6var** *Hina* (= *marama*) *tea atuaatua Korokoro; IKA Mango Ari ... Hina: taha ha, taha ha.* '(It is) the white moon of the (moon) goddess of the month *Koro* (*Korokoro* literally; December chiefly); (it is the star) FISH *Mango ari ...* (It is) the moon: four (nights) have passed, (then) four (nights) have passed.'

In all these texts the observations of the star Spica (or of the whole Virgo constellation) in December are registered. In the third record the appearances of the moon on the fourth and eighth nights/days of the lunar calendar are described (see table 1).

It is known that both terms, Tahitian *Mari-ua* and Maori *Mari-ao*, denote Spica (Best 1922: 32). We now see that the compound term *Mari* = *Ma-ari* (*ari* = bright). Other components of these names describe the motion of the star and its brightness: cf. Tahitian *ua* 'to grow' and Maori *ao* 'bright.' Besides, the Maori name *Whiti-kaupeka* also denotes Spica (Best 1922: 31). In this connection, cf. Maori *whakakau* 'to rise, of heavenly bodies.' The image of the shark/seal (*mango*, *pakia*) as the Spica symbolism could be a designation of the dry summer season, cf. Maori *pakapaka* 'dry,' Rapanui *paka* 'ditto,' Hawaiian *pa'a* 'burnt,' Marquesan *paka* 'dry; burnt,' Mangarevan *pakapaka* 'burnt up,' Maori *whakapaki* 'to dry by fire' and *pakipaki* 'to preserve by drying.'

### The Typological Parallel in the Record on a Babylonian Cuneiform Tablet

Babylonian priest-astronomers stared permanently at the moon far from Easter Island and many centuries earlier. They calculated dates of the first visibility of the new moon (van der Waerden 1974: 242).

### Again on Easter Island: Spica Mentioned in the Local Folklore

In a chant this star was called *Maho* [= *Mango* 'The Shark']-*Rangi* (= in the sky) (Campbell 1971; interpretation in Rjabchikov 1996a: 22):

| | |
|---|---|
| … | … |
| *Ra te na.* | The sun is hidden. |
| *Ko He Ho, ko Maho-Rangi,* | The stars in the sky (are seen:) *He-Hoa* [the star *He* on the eclip- |
| *ko Rangi-Hetuu.* | tic] (and) the Shark [*Maho* = *Mango*] of the sky. |
| *Kohukohu Renga-Mitimiti.* | (It is) the solar eclipse. |
| *Ko te Nuahine Huri.* | (It is) the Black (*Huri* = *Uri*) Old Woman. |
| *Taua; a tae reka.* | (It is) the egg (*toua*); (it is) not good. |
| … | … |

It is the description of a solar eclipse that took place during the annual bird-man festival in the month Hora-nui (September for the major part). I suppose that it is a mythological representation of the total solar eclipse in September 16, 1773 A.D. The star *He* could be either Castor (α Geminorum) or Pollux (β Geminorum) (Rjabchikov 2001: 218-219). The sun was called the epithets *Renga* (the Yellow Colour) and *Mitimiti* (Dried) (Rjabchikov 1998). The invisible new moon was denoted as the Black Old Woman (cf. the wordplay: Old Rapanui *Hina* 'the moon goddess').

The Old Rapanui expression *hungahunga Maho-Rangi* figuratively means 'the god *Makemake* conferred health' (Campbell 1999: 310). In fact, these words denote the period of invisibility of the star Spica in October and November before its heliacal rising. Well, the abundance of eggs of sooty terns (*manutara*) as an important food resource was therefore marked by the disappearance of Spica.

### Spica in the Sky above Rapanui in December: A New Record

Consider this inscription on the Small St. Petersburg tablet, see figure 4.

1(Qv 3): 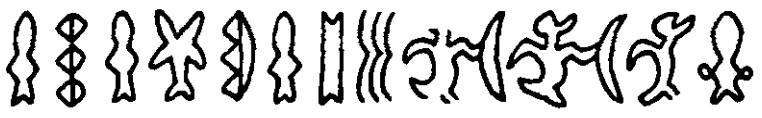

Figure 4.

1 (Qv 3): **73 17-73 11 18-73 4-33 19 3 19 3 19 73** *He tehe Maho, tehe atua ki hina, ki hina, ki He.* 'The star Spica is moving, the deity is moving to the abundant (great) moon, to the star *He*.'

The star *He* (Castor or Pollux) was once high and the moon was low in the sky; then the star *Maho*/*Mango* (Spica) rose. As an example, one can compute such data: Castor: rising: 20:35 on 24 December, 1773 A.D.; Pollux: rising: 20:35 on 24 December, 1773 A.D.; Spica: rising: 00:51 on 25 December, 1773 A.D. (the same night); Moon: setting: 03:02 on 25 December, 1773 A.D. (the same night).

Both Maori terms, *Whakaahu* and *Whakaahu-te-ra,* denote Castor or Pollux (Best 1922: 51), cf. Maori *rā* 'the sun.' It is a key to the meaning of the Rapanui name of the star *He*: cf. Hawaiian *ahu* 'to gather; to collect' and Maori *hea* 'multitude, majority.' Thus, I think that the star *He(a)* was a herald of the high and hot sun in the indigenous beliefs.

### Appendix: A Rat as a Solar Image in the Rapanui Mythology

Let us consider a fragment on the Small Washington tablet (Ra 3-4), see figure 5:

1(Ra 3-4): 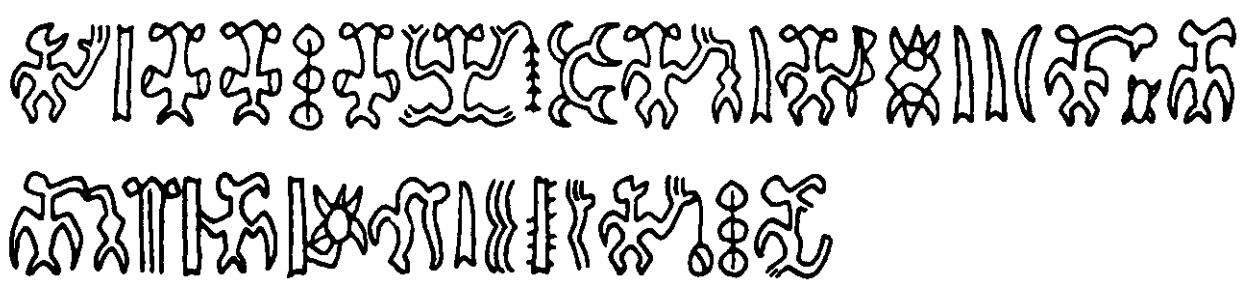

Figure 5.

1 (Ra 3-4): **6-4 72var 72var 17 72var 69-24 8 6 54 5 6-113 7-7 5-5 3 62-28 44 44-54 26-4 44 4 7 44b 5-33 4 15 6 103 17 19** *Hatu manu, manu, te manu Moko-ai; Matua, Hakai atua Hova Tuu Pipiri Atuaatua hina tonga taha. Takai Matua-Taha, atua Tuu tua atuaatua roa; ha pea te kioe.* ‘The scouts appeared at the place Moko (near the Anakena bay); (then king) *Hotu-Matua* (and his wife) *Vakai* [having some features of the goddess *Hina*; cf. Marquesan *hakai* ‘to suckle’] (swam) during the season (*atua Hova*) (associated with) the star Canopus, when the winter turned. (King) *Matua-Taha* (= *Hotu-Matua*) (and the tribal union) *Tuu* in the western part (*tua*) (of the island) were united; (this king) as the Rat gave abundance.’

This text coincides in the general with the brief *rongorongo* records incised on a panel in the royal residence Anakena (Rjabchikov 1996b). It is known that a rat was an incarnation of king *Hotu-Matua* (Barthel 1978: 147). On the other hand, the activity of this king reflected some facets of the local sun religion (Rjabchikov 1995).

## Conclusions

Thus, we go through the secrets of Easter Island to more knowledge about the local archaeoastronomy. The Mataveri calendar has become more understandable for us. A new portion of the *rongorongo* records has been decoded, too.